# Think outside the search box: A comparative study of visual and form-based query builders

Tanja Svarre, Department of Communication & Psychology, Aalborg University, Rendsburggade 14, DK-9000 Aalborg tanjasj@ikp.aau.dk

Tony Russell-Rose, Goldsmiths, University of London, 25 St James's, London, England, SE14 6AD T.Russell-Rose@gold.ac.uk

## Abstract

Knowledge workers such as healthcare information professionals, legal researchers, and librarians need to create and execute search strategies that are comprehensive, transparent, and reproducible. The traditional solution is to use proprietary query building tools provided by literature database vendors. In the majority of cases, these query builders are designed using a form-based paradigm that requires the user to enter keywords and ontology terms on a line-by-line basis and then combine them using Boolean operators. However, recent years have witnessed significant changes in human-computer interaction technologies and users can now engage with online information systems using a variety of novel data visualisation techniques. In this paper, we evaluate a new approach to query building in which users express concepts as objects on a visual canvas and compare this with traditional form-based query building in a lab-based user study. The results demonstrate the potential of visual interfaces to mitigate some of the shortcomings associated with form-based interfaces and encourage more exploratory search behaviour. They also demonstrate the value of having a temporary 'scratch' space in query formulation. In addition, the findings highlight an ongoing need for transparency and reproducibility in professional search and raise further questions around how these properties may best be supported.

## 1. Introduction

Knowledge workers such as healthcare information professionals, legal researchers, and librarians need to create and execute search strategies that are comprehensive, transparent, and reproducible. Healthcare information professionals, for example, undertake systematic searches of scholarly literature sources to identify the best available evidence to inform health and social care policy. Likewise, legal researchers undertake systematic searches to gather evidence to conclusively answer a legal question or support a particular legal position or argument. And librarians undertake all manner of structured search tasks as part of their service delivery to end users and as part of their training and enablement activities for the communities they serve.

However, systematic literature reviews can take years to complete (1), and new research findings may be published in the interim, leading to a lack of currency and potential for inaccuracy (2). Likewise, legal research can be compromised by ineffective or incomplete search, leading to potential loss of legal cases. And librarians routinely formulate complex Boolean expressions, sometimes consisting of hundreds of search terms, leading to significant challenges in maintenance, editing, and debugging (3).

What these professions have in common is a need to develop search strategies that are comprehensive, transparent, and reproducible (4). This typically requires the application of specialist skills in search techniques and often requires specialist expertise in the subject matter in question. This contrasts sharply with other kinds of search tasks, such as casual search (5) and web search (6) which are typically performed on a discretionary basis by users who are not generally expert searchers or subject matter experts.

[insert Figure 1.]

*Figure 1: A traditional form-based query builder (PubMed).*

The traditional solution to the challenge of systematic searching is to use form-based query builders such as that in Figure 1. This approach has its origins in the earliest literature databases, when retrieval was mediated by command line terminals accessing subscription-based services. This required the use of cryptic commands formulated using a proprietary database syntax. The output of the query formulation process would be a sequence of Boolean expressions consisting of keywords, operators and ontology terms which could then be combined using their line numbers to form a composite artefact known as a search strategy (Figure 2).

[insert Figure 2.]

*Figure 2: An example Boolean search strategy (Leeflang et al., 2015).*

However, recent years have witnessed significant changes in human-computer interaction technologies and capabilities in both the home and the workplace. Users now interact with all manner of devices through touch, voice and other modalities, and engage with online information systems using a variety of novel data visualisation techniques. In parallel with this, there have been various studies investigating the use of data visualisation and related techniques to represent structured information needs using a variety of alternative formalisms (see Section 2). However, very few of these alternative approaches have gained any meaningful traction within the information professional or database vendor community.

In this paper, we investigate one such alternative, in which the form-based approach is replaced by a visual canvas on which query elements can be manipulated to represent structured information needs. We compare this approach to a conventional form-based query builder using a number of expert search tasks performed by specialist professional searchers. We evaluate the relative performance of each approach in a controlled test setting (7) and report our findings using a variety of quantitative and qualitative metrics.

# 2. Background

## 2.1 Current approaches

The practice of using Boolean strings for structured searching has served as the default approach since the very first literature databases, and is adopted as the de facto standard by the majority of information databases, vendors and their users. However, despite their ubiquity, using Boolean strings to express complex information needs suffers from a number of shortcomings.

First, Boolean strings are an ineffective vehicle for communicating structure. The use of parentheses as delimiters may be commonplace in programming languages and other machine readable media, but when intended for human interpretation they rely on additional physical cues such as indentation. In the absence of such visual signals, parentheses can become lost in an undifferentiated sequence of alphanumeric characters, and trying to interpret the meaning and structure of such expressions can be overwhelming (8).

Second, Boolean strings do not scale well. As users add terms to a Boolean string, it grows monotonically in length. This may be manageable for a handful of terms, but as the number grows to double figures and beyond, transparency degrades. A common solution to the analogous problem in software development is to offer some form of abstraction, so that lower-level details become progressively hidden, revealing higher-level structure. But Boolean strings in their native textual form offer no such facility (9).

Third, Boolean strings are error-prone: even if the query builder automatically checks syntax, it is still possible to place parentheses incorrectly, which can inadvertently change the semantics of the whole expression. This gives rise to errors whose effects may not become apparent until long after the initial search is completed (10).

In form-based query builders, users are expected to build search strategies incrementally as a set of discrete expressions that are referenced by line number and then combined using various operators. This type of approach offers the benefit that strategies can be built using techniques such as successive fractions or building blocks (11). It also allows the searcher to review the number of results returned at each step and to refine their expressions accordingly.

However, errors and inefficiencies often compromise their output. In their study of sixty-three MEDLINE strategies, Sampson and McGowan detected at least one error in over 90 percent of these, including spelling errors, truncation errors, logical operator error, incorrect query line references, and redundancy without rationale (12).

Many information professionals use word-processing tools such as Google Docs or Microsoft Word as a platform for developing search strategies (13). However, there are several reasons why document-centric tools are not suitable for search strategy formulation:

- Auto-correction can undermine truncation and corrupt truncated formats
- Spell checking can obfuscate important differences between regional linguistic variations (e.g. British and American English) and can create unwanted duplicates
- Auto-conversion of straight quotations ("") into curly quotations ("") can cause errors in platforms such as Ovid SP
- Copying and pasting text fragments between different word-processing tools can lead to loss of non-print characters.

This combination of factors suggests that the practice of formulating search strategies as text strings manipulated via form based query builders compromises their ability to function as transparent, scalable, reproducible artefacts.

## 2.2 Alternative approaches

The application of data visualisation to search-query formulation can offer significant benefits, such as fewer zero-hit queries, improved query comprehension, and better support for exploration of an unfamiliar database (14). Anick et al (15) is an early example of such an approach. They developed a system that could parse natural language queries and represent them as movable tiles on a two-dimensional canvas. The user could rearrange the tiles to reformulate the expression and to activate or deactivate alternative elements to modify the query. In addition, the system offered support for integration with thesauri, and it displayed the number of hits in the lower left corner of each tile.

In subsequent work, Fishkin and Stone (16) investigated applying direct manipulation techniques to database query formulation using a system of "lenses" to refine and filter the data. Users could combine lenses by stacking them and applying a suitable operator or combine them to create compound lenses, supporting the encapsulation of complex queries. Jones (17) proposed an influential approach in which concepts are expressed using a Venn diagram notation combined with integrated query result previews. Users could formulate queries by overlapping objects within the workspace to create intersections and disjunctions, and they could select subsets to achieve a further refined set of results.

Yi et al. developed a system based around a "dust and magnet" (18) metaphor, in which users could represent dimensions of interest within the data as magnets on a visual canvas. The effect of the "magnetic forces" on individual "data particles" reflected the relationships between points in the data. Nitsche and Nürnberger (19) developed a system based around a radial interface in which users could integrate and manipulate queries and results. The concept used a pseudo-desktop metaphor in which objects of interest clustered toward the centre. Query objects could be entered directly onto this canvas, and their proximity to the centre and to other objects was a relevance cue, influencing the selection and position of search results.

More recently, Scells and Zuccon developed searchrefiner, an open source tool for formulating, visualising, and understanding Boolean queries (20). This tool allows researchers to perform tasks such as using validation citations to ensure queries are retrieving a minimum set of known relevant citations, and editing Boolean queries by dragging and dropping clauses in a structured editor based on a hierarchical tree metaphor. In addition, the tool allows researchers to visualise why the queries they formulate retrieve citations, and to understand how to refine queries into more effective ones. Likewise, the practitioner community have also made contributions to support structured query formulation, notably the Search Whiteboard[1] developed by David Newman for searching healthcare databases, and Boolio[2] developed by Wrenford Thaffe for sourcing social media profiles for recruitment and marketing purposes. Both of these tools use a 2D grid to represent disjunctions in one dimension and conjunctions in the other (implemented using an Excel spreadsheet in the former case and a web application in the latter).

## 2.3 Research questions

In this paper, we investigate the following research questions:

1. How does the use of a visual vs. form-based interface influence the **process** of structured searching?
2. How does the use of a visual vs. form-based interface influence the **user experience** of structured searching?
3. How does prior experience of structured searching influence participants' **attitudes and expectations** regarding search interfaces?

# 3. Materials and methods

In our experimental setup we investigate a form-based approach and a visual approach and compare them in a controlled test setting using a variety of quantitative and qualitative metrics. We used PubMed's advanced search query builder as an instance of a conventional form-based interface, and 2Dsearch[3] as an instance of a visual interface. Both interfaces were used to query the same database (PubMed), with the latter acting as a replacement for PubMed's native (form-based) query builder. We compare the two approaches using a number of expert search tasks performed by specialist professional searchers in a controlled test setting (7). We evaluate the relative performance of each approach and report our findings using a variety of quantitative and qualitative metrics. In the remainder of the paper we will refer to these two system types as the *form-based* and *visual* interfaces respectively.

## 3.1 Participants

The participant group consisted of 14 experienced research librarians (3 men and 11 women), recruited from two Danish university libraries. Their average age was 49 years.

---

[1] https://exeterhealth.libguides.com/searching/Resources
[2] https://www.scoperac.com/products
[3] https://www.2dsearch.com/

## 3.2 Tasks

The participants explored the two interfaces by means of one training task and four simulated work tasks (21). The purpose of the training task was to let the participants familiarise themselves with the experimental system without recording their activity. The tasks were designed to exercise the participants' technical and professional search expertise, using healthcare and life sciences as the chosen subject domain. Within this domain the tasks represented a library and information science perspective to enable the participants to apply their topical knowledge, which have been shown to be important in simulated work tasks for professionals (22). All tasks were selected to include at least three discrete facets to the information need in order to require the use of structured search methodology. Figure 3 illustrates an example task. All tasks were pilot tested with two additional participants prior to the data collection to ensure relevance and understanding for potential test participants. Based on the pilot test results, minor corrections were made.

---

Task A

**Title: Patient information seeking**

Description: Find studies of how patients look for online information on their cancer diagnosis

Narrative: Relevant documents will contain a description of purposes, information sources, approaches or the like.

---

Figure 3: One of the four search tasks.

The sequence of tasks was rotated for each session to minimise any order effects. After each task, participants completed a short questionnaire to identify any other factors affecting their performance, such as prior knowledge of the task domain. In addition, they were asked to assess the difficulty of the task and provide responses to various user experience metrics (23) regarding the search process and the results of the search process. A total of 55 task assessments were collected.

## 3.3 Procedure and data collection

The data collection took place in Autumn 2020. The data collection consisted of three stages: a pre-session survey completed prior to the session, the test session itself, and a post-session interview.

The pre-survey served several purposes. First, it was used to collect the participants' consent for participation in the study. Second, the participants were asked questions about their search preferences, such as using simple vs advanced search, previous use of Boolean expressions, and confidence in using Boolean operators. Demographic questions concerning gender and age were also included.

At the beginning of each session, participants signed an informed consent form on paper and were introduced to the test procedure. The training task was then administered, followed by the four simulated work tasks that were rotated between participants for both tasks and systems (24). After completing all the tasks, a semi-structured interview was performed, in which participants were asked about their experiences in using both the form-based and the visual interface.

After finishing all test sessions, the interviews were transcribed, translated, and analysed. The search log and survey elements were analysed using statistical analysis, while the interviews were coded in themes using Nvivo for a systematic analysis of the findings. Queries were analysed at both session and query level.

# 4. Results

The pre-survey revealed that study participants had a general preference for advanced search (10 participants) over simple search (4 participants). All of them felt secure or very secure in using Boolean operators, which is also reflected in their subsequent use of such operators. 8 out of 14 participants stated that they used Boolean AND on a daily basis, 9 used Boolean OR on a daily basis or several times a week, and 11 used phrase search daily or several times a week. Boolean NOT was less common with 13 participants indicating that it was used once a month or less. The overall picture is that the participants were experienced users with a preference for advanced search, entailing frequent use of boolean operators. This corresponds with the findings made by Russell-Rose & Chamberlain (25). In a survey of medical librarians they found that boolean logic was the most important functionality of query formulation.

## 4.1 Tasks

Participants were asked to provide feedback on each task and any background knowledge they had on the specific topic. The results are shown in table 1. Overall, the participants did not claim to have much prior knowledge of the task domain (mean=1.6). From the general assessments it appears that despite the feeling of limited knowledge about a particular task, the participants find the tasks clear (mean=1.49), more simple than complex (mean=2.47), more easy than difficult (mean=2.76), and more exciting than boring (mean=3.69). These other task assessments could be a reflection of the fact that in the assessment of prior knowledge, the participants may not have had extensive knowledge of the medical context. However, they were still able to complete each task by applying their general knowledge of library and information science topics in their capacity as trained librarians. When focusing on specific tasks Table 1 shows that there was some variation between the assessments but no significant correlations were found between task assessments and the number of reformulations made at session level.

|  | Task A (N=14) | Task B (N=13) | Task C (N=14) | Task D (N=14) | Total (N=55) |
|---|---|---|---|---|---|
| Prior knowledge of task topic | 1.71 | 1.54 | 1.64 | 1.50 | 1.60 |

| (1-5, 1 is low) | | | | | |
|---|---|---|---|---|---|
| 1 - Clear<br>5 - Unclear | 1.43 | 1.08 | 1.57 | 1.86 | 1.49 |
| 1 - Simple<br>5 - Complex | 2.29 | 1.85 | 2.93 | 2.79 | 2.47 |
| 1 - Boring<br>5 - Exciting | 3.93 | 3.85 | 3.43 | 3.57 | 3.69 |
| 1 - Easy<br>5 - Difficult | 2.57 | 2.38 | 3.07 | 3.00 | 2.76 |

Table 1: Assessments of simulated work tasks
Legend: all 5 scales were measured on a 1-5 scale.

## 4.2 Queries

Across the 55 sessions a total of 203 queries were generated. The queries carried out had a mean of 2.22 facets out of three possible, and a mean of 4.56 terms. Table 2 illustrates the variation across the controlled tasks of the evaluation.

| | Task A | Task B | Task C | Task D | Total | Total st.dev. | Total min. | Total max. |
|---|---|---|---|---|---|---|---|---|
| mean no of terms | 4.11 | 3.37 | 5.02 | 5.22 | 4.56 | 2.93 | 1 | 13 |
| mean no of facets | 2.32 | 2.37 | 2.21 | 2.11 | 2.22 | 0.824 | 1 | 3 |

Table 2: mean number of terms and facets in total and for specific tasks

An independent samples t-test showed the differences in terms, facets, and overlap with experts across the two interfaces. As seen in Table 3 participants using the visual interface generated queries with a significantly higher number of facets and a significantly higher number of terms. Several themes in the post-interviews may provide an explanation for these results. One theme is that trained users tend to form queries for individual synonyms first, for example by a look-up in a controlled vocabulary in the form-based interface, and then subsequently combine them into facets which would generate a greater number of shorter queries using this particular interface. To illustrate: "*[...] it is because I prefer to conduct the single term query before I form the block…*" (TP12). On the other hand, several participants expressed how their usual process of planning a complex query with pen and paper before approaching a database can now be conducted directly using the visual interface. For example TP12 states: "*Well, it is easier in [the visual interface], because of those blocks… I would do the same thing if planning a query in real life. I would actually draw squares on a piece of paper*". And TP2 adds: "*Normally, I would have a piece of paper, where I would make little boxes with the blocks. And perhaps that can be merged, because

*that is actually what is done in [the visual interface]. [...] You can save some time by doing it directly in the interface*".

In addition, several participants explored the functionalities in the visual interface, which had the effect of increasing the number of terms and facets in the visual interface. As expressed by TP13: "*I like the visual part. [...] the moving things around, combining them, back and forth, and so on. [...] It is pleasant to use. [...] You sit and play with it. It is kind of a building block technique, you could say. [...] It is sort of fun. You get captured by it.*"

|  | Form-based (N=105) | Visual (N=98) |
|---|---|---|
| Mean number of terms | 3.29*** | 5.92*** |
| Mean number of facets | 2.03*** | 2.43*** |
| Correlation between terms and facets (Pearson's R) | 0.759*** | 0.358*** |

Table 3: Differences between form-based and visual interfaces at query level
Significance: *= > .1, **= > .05, ***= >.001

The results show that significantly more terms and facets were used in the visual interface. In both interfaces, there was a significant correlation between the number of terms used and the number of facets used. This correlation was stronger for the form-based interface, which may reflect a more homogeneous approach adopted by the experienced searchers when using an interface with which they were already familiar. By contrast, the visual interface attracted more variation in its use, possibly due to the novelty issues as suggested above. To illustrate, TP6 states that "*...especially when this is not my usual topic area, the [visual interface] provides some possibilities to just try things out and see what happens…*", implying that the visual interface acted as a kind of experimental 'scratch space' for the participant, because she solving a task outside her known area of expertise. Further, this comment reflects how a lack of insights into the task topic plays a role in how the query is formed.

## 4.3 Sessions

In performing the four simulated work tasks across the two systems, 55 sessions were carried out, 29 sessions with the form-based interface and 26 with the visual interface. Table 4 summarises the differences between the two approaches. As can be seen, the users made slightly more query reformulations using the form-based interface (3.24), compared to visual (2.96). However, there was a larger variance associated with the visual interface with a standard deviation of 3.08 compared to 2.77, suggesting that there was a greater degree of variation in the actual use of the visual interface. Several elements in the post interviews may help explain these effects. Some of the librarians explain how they identify terms for one facet, or even one synonym for a facet, of the search task at the time and then combine them subsequently when using the form-based interface: *"..so if I have three synonyms in [a form-based interface] that I would combine with OR, then I would make a query for each synonym separately and then combine them subsequently… because then I can quickly see*

*which terms work and which don't…" (participant 12)*. And: *"[in traditional search] you make a very advanced search [history] and suddenly you have 100 queries on the line. Then I always write down which ones to AND in the end" (participant 2)*

To some participants the search history is important in order to identify the task facets and subsequently combine them in order for the final query to represent the complete search. As reformulations are equivalent to the number of queries for a task, this way of controlling the outputs in PubMed would add to an increased number of reformulations. Conversely, other participants state that the canvas of the visual interface allows them to generate several facets in parallel, thus reducing the need for (and number of) explicit reformulations: *"I can quickly search for what I want, because I throw in three concepts in three different blocks and see what it gives me." (TP3)*

However, the larger degree of variance associated with the visual interface may be explained by some participants approaching the interface in a more playful manner to explore its potential, which would generate more queries in a session. Others expressed that they tend to make a more complete query in the visual interface before checking the results of the query, which would lead to fewer queries in a session. TP12 states that: "*Then I would sit in [the visual interface] and do something and then in the end [look at the results]. So it is somehow more conclusive, before I check the results.*" And later she continues on the visual interface: "*I jump faster to the conclusion that these three blocks are relevant and good together, than when I use [the form-based interface].*" This would ultimately lead to fewer reformulations in the visual interface.

|  | Form-based (mean) | Form-based (st.dev) | Visual (mean) | Visual (st.dev) |
|---|---|---|---|---|
| Mean number of reformulations (N=48) | 3.24 | 2.77 | 2.96 | 3.08 |
| Mean assessments of search process: |  |  |  |  |
| Simple >< complex (N=55) | 2.45 | 1.21 | 2.35 | 1.06 |
| Pleasant >< unpleasant (N=55) | 2.28 | 1.07 | 2.00 | 1.06 |
| Difficult >< easy (N=55) | 3.17 | 1.14 | 3.38 | 1.17 |
| Relaxing >< exhausting (N=55) | 2.59 | 0.91 | 2.23 | 0.86 |

Table 4: Differences in form-based and visual interfaces at session level

The two systems also differed in terms of the participants' assessment of the search process in the two systems. In general the visual interface was associated with a better search experience in terms of the simplicity of search (2.35 vs 2.45), being more pleasant (2.00 vs 2.28), easier to use (3.38 vs 3.17), and less exhaustive to use (2.23 vs 2.59).

The interview explains some of the overall evaluations. One aspect concerns the ease of adjusting a query, if it turns out not to provide a desired outcome. TP2 reflects on this: "*It is easier to adjust a query [in the visual interface] because the box is just there. [in the form-based interface] I have to edit a query I made earlier, and if I combined it with something else, I need to go further back, because I can't add it subsequently.*" This is supported by TP9, who mentions that "*I get a better overview [in the visual interface]. because it is easier to pull things apart and draw things into the query again. It is too much trouble in form-based interfaces in general.*" The flexibility of the visual interface is also mentioned by TP8: "*At one point in [the form-based interface] I suddenly had zero results, and I wasn't sure if I had misspelled something, or if I had put it in the wrong place, or if there just weren't any results. And to compare, I think it was easier in [the visual interface] to add or remove elements to see the influence.*" Also the lack of overview in form-based interfaces was commented on by some participants. To illustrate: "*It is clear that you very quickly lose track in [form-based interfaces], when doing deep and large blocks. It is easy to lose the big picture of what I have been doing*" (TP3). In this respect, the traditional form-based interface forces the user to think sequentially, treating each element as a command within a larger sequence that must then be combined with other commands according to their respective positions within the sequence. By contrast, the visual interface allows users to think in parallel, experimenting with different combinations regardless of the order in which they have been articulated.

Despite the positive assessments of the visual interface, several participants articulate the familiarity and security of the well-known form-based interface. According to TP12: "*This way of searching is what I am used to, so it is a habit that is difficult to change.*"

The form-based interfaces also support transparency in the queries carried out by the expert searchers because of their familiarity with this way of searching. One participant elaborates on how she performs queries in the visual interface, but uses the form-based interface to interpret the query to find out what was actually done in the visual query: "*I am more certain about the results in the form-based interface, because I can figure out the results there. [...] I got used to checking the visual query in the form-based interface to validate what I actually searched for. It is very important as a control feature.*" (TP14). Another participant agrees and requests transparency in the form of a clear view of the syntax behind the visual interface: "*I can't see the syntax behind the query*" (TP3).

Another participant felt a lack of experience in how specific operators were used in the visual interface, which led to a reflection about how operators are used in form-based interfaces to make small adjustments in queries: "*I felt a bigger flexibility in [the form-based interface]. I wanted to use some parentheses in [the visual interface], but I gave up on it, because I didn't know how to do it*" (TP2).

To sum up, this analysis has shown that use of the visual interface generates more complex queries with a greater number of facets and terms per query. The visual approach helps the expert searcher adjust queries and understand the search process with a better overview of the steps taken. However, the expert searchers are also accustomed to understanding transparency, control and validation in terms of their representation as boolean logic within form-based interfaces, and this convention is missed in the visual interface tested in the current study.

# 5. Discussion

In this paper, we compare a conventional form-based query builder with a visual approach in which query elements could be manipulated as objects on a visual canvas. We found that visual interfaces can mitigate some of the shortcomings associated with form-based interfaces and encourage more exploratory search behaviour through the provision of novel interactions such as a temporary 'scratch' space for query formulation. However, the results also highlight the enduring importance of qualities such as transparency and reproducibility in professional search, and this raises important issues around how these properties may best be supported. It is to these issues that we now turn our attention.

First, let us consider the formalism itself, i.e. the convention of using Boolean strings and line numbers to represent a composite logical expression. If the purpose of the formalism is to provide a reproducible mechanism for representing structured information needs, is it appropriate to rely on something as arbitrary and ephemeral as a line number? This is the conceptual equivalent of the GOTO statement used in first generation BASIC, which is an approach that was discredited several decades ago (26). By contrast, and taking inspiration from the discipline of software engineering, we might expect a well-designed query language to support properties such as:

- Encapsulation: i.e. the bundling of data with the methods that operate on that data. This is not supported natively by Boolean strings, but the visual approach shows that terms and associated operators can be packaged into a single component and represented physically within the interface.
- Abstraction: i.e. the creation of abstract concepts by mirroring common features or attributes of various non-abstract objects. Again, this is not supported natively by Boolean strings, but the visual approach demonstrates how repeatable constructions such as the 'problem', 'intervention', 'comparison' and 'outcome' elements of a PICO search can be abstracted out as visual templates (27).

Secondly, as Figure 2 illustrates, the output of form-based query builders is a set of *procedural* commands that are combined to express the *declarative* semantics of an information need. This raises the question: if the goal is to express declarative semantics, then why force the user to think procedurally? This may be due in part simply to the power of convention, in that the command-line approach originates from an era when searchers could issue only text-based commands to remote—and expensive—subscription-based resources. In that historical context, it may have made sense to prioritise economy over usability. However, the continued adoption of this convention in contemporary search interfaces may owe more to inertia than any inherent design virtue.

The findings of this study suggest that the development of alternatives to Boolean strings and form-based query builders may not just be possible, but also desirable. However, we should also note a number of limitations. In what follows we review these limitations and identify their potential impact on the specific research questions (shown in parentheses).

First, there is the issue of prior knowledge of the search tooling. Familiarity with conventional interfaces could influence both the performance and preferences of participants. In particular, participants may naturally gravitate toward what they know best, and this could have an impact on the results. (RQ1, RQ2). Second, there is the issue of prior knowledge of the tasks: participants had limited knowledge of the medical domain (although most were still sufficiently confident that they could complete the tasks). The outcomes may have been different had the participants had greater familiarity with the subject matter (RQ1). Third, there is the issue of prior expertise in structured searching: The participants were all information professionals who were highly skilled in the concept and execution of structured searching. The outcome may have been different had the participants been novices or individuals with limited experience in structured searching (RQ3). Fourthly, there is the fact that this study used just one exemplar of each interface type. Evidently, it is possible to imagine other types of visual interface, which offer a similar set of interactions but via a different design execution. Although a number of the findings are statistically significant, it would not be appropriate to generalise too far based on a single instance of each interface type (RQ2).

## 5.1 Future work

The limitations identified earlier suggest a number of avenues for need for future work. First, this study was carried out in a controlled experimental setting, and further studies would be needed to validate our understanding of searching with visual interfaces within a naturalistic setting. Second, our methodology was to document interaction by means of surveys, screen recordings, and follow up interviews. The use of alternative methods such as eye tracking could have provided additional insights into the user interactions taking place, and future studies should consider this methodology. Third, this study focused on one instance each of a form-based and a visual interface, and further work is needed to determine the extent to which these findings generalise to other examples and on the impact of different kinds of visual interface on the search process. Finally, this study has focused on investigating the use of visual interfaces by search professionals. Further work is needed to understand the impact on different user types such as recruiters and other professionals, and also on non-professional searchers and discretionary users who may exhibit different attitudes, expectations and approaches.

# 6. Summary and conclusions

In this paper, we have investigated a form-based interface and a visual interface to search strategy formulation using a variety of quantitative and qualitative metrics. We compare them on a number of expert search tasks performed by specialist professional searchers in a controlled test setting. We now draw conclusions in relation to our original research questions. First of all we wanted to understand how the use of a visual vs. form-based query builder influences the process of structured searching. We found that the search professionals used significantly more terms and facets in the visual interface, as they used it to structure their search directly in the interface rather than using pen and paper on the side, which in many cases was their usual approach. The number of terms and facets was reduced in the form-based interface, as more single facets were developed one by one and subsequently combined. On this basis it can be concluded that the visual approach to structured searching supports a more holistic approach, where complex queries are developed as a whole instead of being broken down into facets that are developed on an individual basis.

Secondly, we aimed to understand how visual vs. form-based query building influences the experience of structured searching. On all the user experience metrics collected during the testing the visual approach was rated better than the form-based approach. The results indicate that there is a great potential in incorporating visual elements in search interfaces, including even those designed for trained professional searchers.

Lastly, we asked the question how prior experience of structured searching influences participants' attitudes and expectations regarding search interfaces. The analysis of results demonstrated how test participants that were fundamentally used to form-based approaches on one hand valued the user experience of the visual approach, but at the same time relied on the familiarity and transparency of form-based interfaces, where it is evident exactly what has been searched for, and how. This finding may also reflect a general lack of experience among participants with visual approaches in the test setting. Another conclusion that can be drawn from this is that in the development of search aimed towards search professionals, transparency of the approach remains a vital criterion to reflect the explicitness of structured, expert searches.